\documentclass{IEEEtran}
\usepackage{graphicx,cite}
\usepackage{epsfig}
\usepackage[cmex10]{amsmath}
\usepackage{amssymb}
\usepackage{array}
\usepackage{fixltx2e}
\usepackage{amsfonts}
\usepackage{lineno}

\begin{document}
\title{Implementation of Distributed Time Exchange Based Cooperative Forwarding}

\author{\IEEEauthorblockN{Muhammad Nazmul Islam\IEEEauthorrefmark{0},
Shantharam Balasubramanian\IEEEauthorrefmark{0},
Narayan B. Mandayam\IEEEauthorrefmark{0}, Ivan Seskar\IEEEauthorrefmark{0} 
 and Sastry Kompella\IEEEauthorrefmark{1}  \\
 }
 \IEEEauthorblockA{\IEEEauthorrefmark{0}
Wireless Information \& Networking Laboratory, Rutgers University \\
mnislam@winlab.rutgers.edu, shantharampsg@gmail.com,
narayan@winlab.rutgers.edu, seskar@winlab.rutgers.edu,
\IEEEauthorrefmark{1} Information Technology Division,
Naval Research Laboratory, Email: sk@ieee.org}}

\maketitle

\begin{abstract}

In this paper, we design and implement time exchange (TE)
based cooperative forwarding where nodes use transmission 
time slots as incentives for relaying.
We focus on distributed joint time slot exchange and relay selection
in the sum goodput maximization of the overall network.
We formulate the design objective as a mixed integer nonlinear programming (MINLP)
problem and provide a polynomial time distributed solution of the MINLP. 
We implement the
designed algorithm in the software defined radio enabled USRP nodes
of the ORBIT indoor wireless testbed. The ORBIT grid is used 
as a global control plane for exchange of control information between the
USRP nodes. Experimental results suggest that TE can significantly
increase the sum goodput of the network. We also demonstrate
the performance of a goodput optimization algorithm
that is proportionally fair.

\end{abstract}


\section{Introduction}

Cooperative forwarding improves the connectivity and 
throughput in wireless networks~\cite{Tse}. However, forwarding 
always incurs some costs, e.g., delay/power, at the
forwarder node. Hence, there have been recent studies on 
resource delegation based incentivized
forwarding~\cite{Zhang, Baochun:a,Ray,Nazmul2,Zhang2,Lindstorm}.
Resource delegation based forwarding allows the sender node to delegate a portion of its 
allotted resource to the forwarder node as an immediate incentive for relaying.
Previous works on incentivized forwarding have focused on resource exchange from a theoretical
perspective and used numerical simulations to justify the
effectiveness of the approach. The main
contribution of our work is to demonstrate the advantages of incentivized forwarding 
using the ORBIT indoor wireless testbed~\cite{ORBIT}.

\begin{figure}[t]
\centering
\includegraphics[scale=0.35]{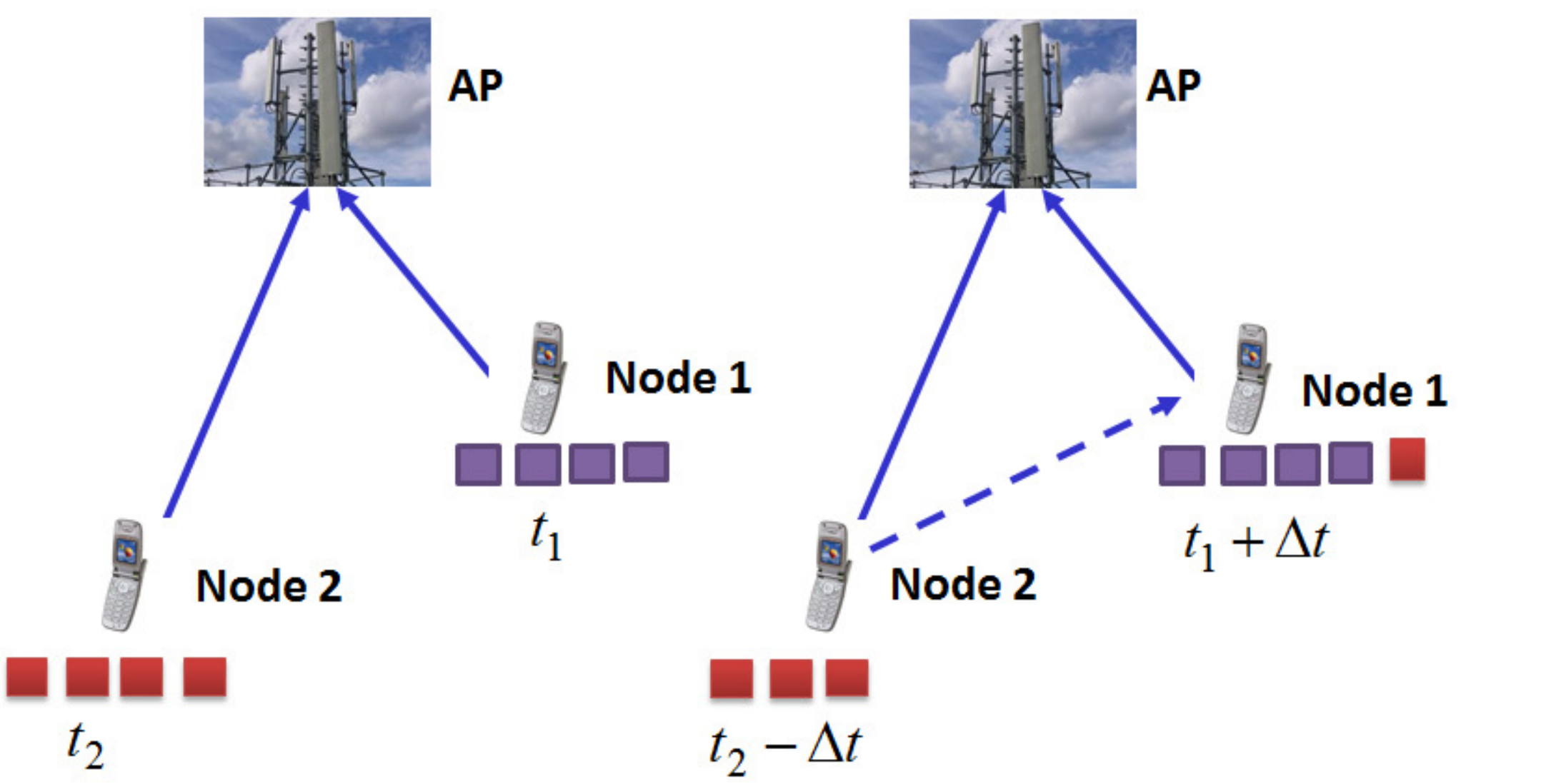}
\caption{The motivating scenario for time exchange based cooperative forwarding}
\centering
\label{fig:TimeExchange}
\end{figure}

We specifically focus on the uplink of an $N$ node time division 
multiple access (TDMA) network where each node 
receives an initial number of time slots and transmits data to the 
base station (BS) through the direct path. In this context, we focus on a two hop
time exchange (TE) based forwarding scheme where the sender node transfers a portion 
of its allotted time slots to the forwarder node as an incentive for relaying. 
The basic idea of TE is illustrated in Fig.~\ref{fig:TimeExchange}.
In this example, node $1$ and $2$ initially receive $4$ time slots and transmit
through the direct path. In TE,
node $1$ performs a half duplex decode and forward (DF) relaying of node $2$'s data
and node $2$ delegates one time slot to node $1$ as an incentive for relaying. 
Node $2$ may attain higher data rate through this cooperation since its
data goes to the BS through two different paths. Node $1$ may also get higher
data rate since it has more transmission time slots. 
The optimal time slot delegation is an important question in this context.

A sender node can delegate time slots to multiple forwarder nodes 
and transmit data to the BS through multiple paths. 
However, the authors of~\cite{Ravi} have shown that for a source-destination pair, 
in the presence of multiple relay nodes, it is sufficient to select the
``best relay node" to achieve full diversity order. 
On the other hand, the assumption of one sender node for one forwarder node
reduces the relay selection complexity. Therefore, we assume one forwarder node for one 
sender node and vice versa in this work. The optimal distributed
sender-forwarder pair selection problem becomes another
important question in this context.

In this work, we address the joint time slot delegation and sender-forwarder pair
selection question in the context of sum goodput maximization of the overall network.
We formulate the joint optimization problem as a mixed integer 
nonlinear programming (MINLP) problem. 
Using our relay selection work of~\cite{Nazmul2},
we show that the distributed solution of the MINLP requires
$O(k^{tot})$ computational complexity and at most $N^2$ message passing
where $k^{tot}$ and $N$ denote the total time slots and nodes in
the network.
The designed algorithm maximizes the sum goodput of the network 
while preserving the local goodput of the individual nodes.

We implement the designed algorithm in the ORBIT indoor wireless
testbed. We use software defined radio enabled USRP~\cite{USRP} nodes of 
the ORBIT~\cite{ORBIT} testbed.
The ORBIT grid serves as the global coordination plane~\cite{Narayan} to
exchange the protocol information between the nodes. 
The data transmission through the air is processed 
using the GNUradio codes~\cite{GNUradio}.

Previous works in resource exchange~\cite{Zhang, Baochun:a,Ray,Nazmul2,Zhang2,Lindstorm}
have focused on developing information theoretical algorithms.
We design the theoretical framework
in accordance with the testbed constraints and
then implement the framework in the ORBIT testbed.
Our theoretical analysis has similarities to the classical maximum weighted scheduling study of~\cite{Ephremides}. However, unlike~\cite{Ephremides}, we consider a network where each node
starts with a fixed number of time slots and then tries to find the
optimal number of time slot transfers.
Our proposed framework can be applied to TDMA based commerical (GSM \& Edge~\cite{Goldsmith}, 802.16 Wireless MAN~\cite{WirelessMAN})
and tactical (Joint Tactical Radio System~\cite{JTRS} and Link16~\cite{Link16}) networks.

This paper is organized as follows. 
Sec.~\ref{sec:Model} and~\ref{sec:Theory} illustrate the
system model and design objective respectively.
We solve the optimization problem in Sec.~\ref{sec:Solution}.
After describing the experimental setup in Sec.~\ref{sec:Setup},
we demonstrate the testbed results in Sec.~\ref{sec:Results}.
We conclude the work in~\ref{sec:Conclusion}.

\begin{figure}[t]
\centering
\includegraphics[scale=0.27]{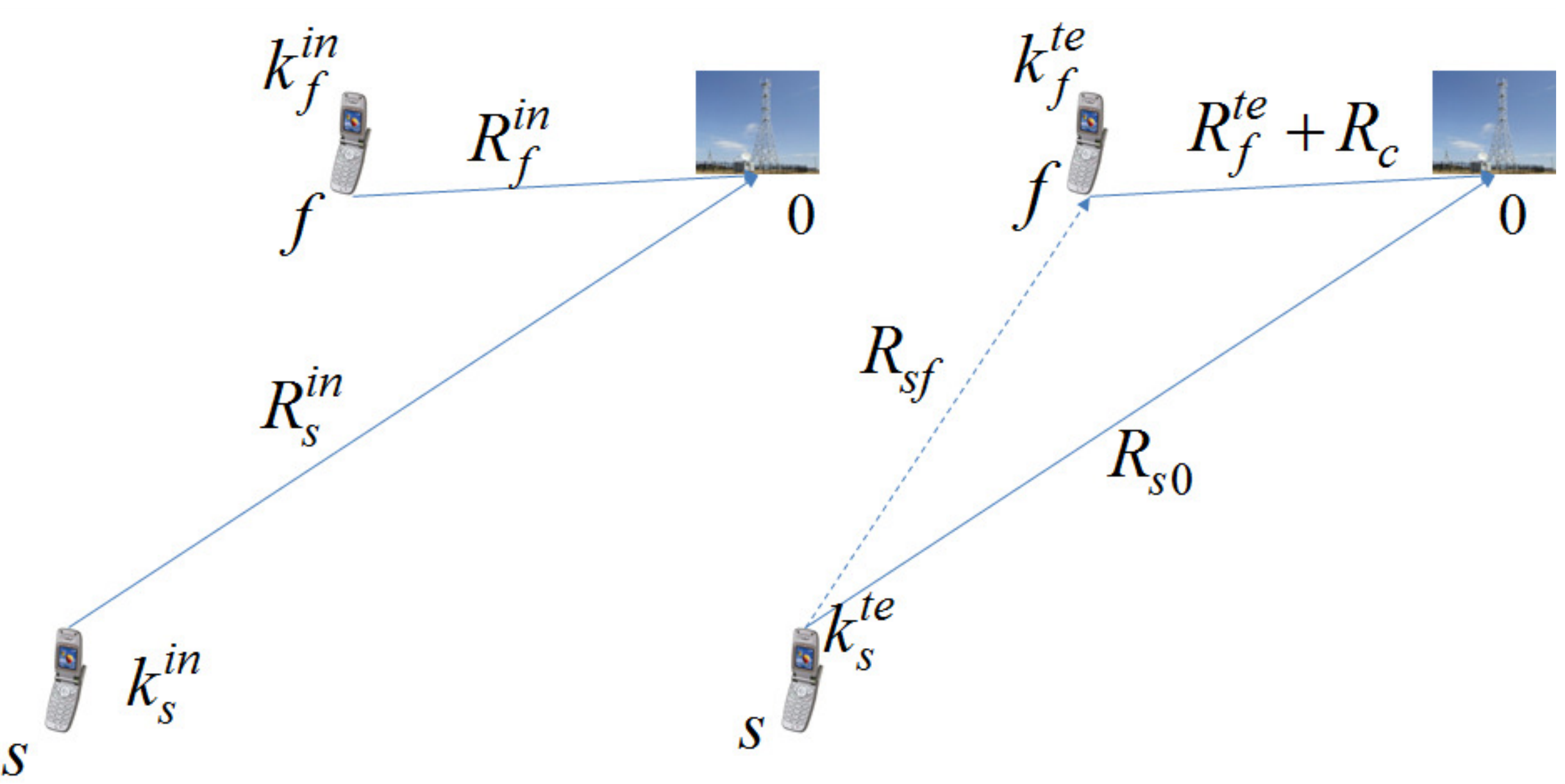}
\caption{Direct Transmission and Time Exchange System Model}
\centering
\label{fig:packet_transmission}
\end{figure}

\section{System Model}   \label{sec:Model}

We consider the uplink of an $N$ node
single cell TDMA network. Let $\mathcal{V} = \{1, \, 2, \, \cdots, \, N\}$
denote the set of $N$ nodes that transmit
data to the BS (node $0$). Each node uses the same bandwidth.
Node $i \in \mathcal{V}$ is initially allotted $k_i^{in}$
time slots per second. Without loss of generality, assume
one packet is transmitted per time slot. 
Each node is assumed to employ a fixed modulation scheme.
This assumption is based on the testbed implementation
constraint and will be further explained in the experimental setup
section. Due to the fixed modulation and total bandwidth
usage, the data transmission rate of each node depends on the 
number of allotted time slots. 

We use packet loss probability as the channel strength indicator. 
Let $Pe_{ij}$ denote the packet loss probability in the $ij$ path.
Define goodput of node $i$ by the number of packets of node $i$
that successfully reach the BS. Node $i$ initially transmits through 
the direct link $i0$. Hence, the initial goodput of node $i$, $R_i^{in}$,
can be found as:
\begin{equation}
R_i^{in} = R_{i0} = k_i^{in} * (1 - Pe_{i0})    \label{eq:Packet_Direct}
\end{equation}
\subsection{Goodput Analysis in TE}

In TE, nodes perform two hop half duplex 
decode and forward (DF) relaying. Let, 
$\mathcal{SF} = \{\mathcal{SF}_1, \cdots, \mathcal{SF}_k\}
=\{(s_1,f_1),(s_2,f_2),\cdots,(s_K,f_K)\}$
denote the sender-forwarder pairing sets, i.e., the forwarder node $f_i$
relays the sender node $s_i$'s data, along with 
transmitting $f_i$'s own data. Let $\mathcal{D} = {d_1, d_2, \cdots, d_L}$ denotes
the direct set, i.e., the set of remaining nodes that transmit data without cooperation.

The left and right hand side of Fig.~\ref{fig:packet_transmission} shows the 
direct transmission and TE model of an
arbitrary sender-forwarder pair $(s,\,f)$.
Node $s$ and $f$ initially receive $k_s^{in}$ and $k_f^{in}$
time slots and obtain $R_s^{in}$ and $R_f^{in}$ goodput
respectively.

\begin{table}
\caption{Summary of used notations} \label{Tab:one}
\begin{center}
\begin{tabular}{|l|l|} \hline
Notation & Meaning \\ \hline
$Pe_{ij}$ & Packer error probability of the $ij$ link \\ \hline
$k_i^{in}$ & Initial transmission time slot of node $i$ \\ \hline
$k_i^{te}$ & Node $i$'s btransmission time slot in BE \\ \hline
$R_i^{in}$ & Initial goodput of node $i$ \\ \hline
$R_i^{te}$ & Node $i$'s goodput in TE \\ \hline
$R_{ij}$ & Goodput in the $ij$ link \\ \hline
$\mathcal{V}$ & Set of $N$ nodes \\ \hline
$\mathcal{D}$ & Set of nodes that transmit without cooperation  \\ \hline
$\mathcal{SF}$ & Set of sender-forwarder pairs \\ \hline
$\mathcal{SF}_i$ & $i^{th}$ sender-forwarder pair $(s_i,f_i)$ \\ \hline
$s$ & Sender node \\ \hline
$f$ & Forwarder Node \\ \hline
$k^{tot}$ & Total time slots in $1$ second \\ \hline
$n_i$ & Number of neighbouring nodes of node $i$  \\ \hline
\end{tabular}
\end{center}
\end{table}

In TE, node $s$ transmits for $k_s^{te}$ time slots. 
During $k_s^{te}$ time slots, node $s$ transmits $k_s^{te}$ packets. 
Among these packets, the BS and node $f$ receive 
$R_{s0}$ and $R_{sf}$ `error-free' packets respectively. 
Following the analysis of~\eqref{eq:Packet_Direct},
\begin{equation}
R_{sf} = k_s^{te} * (1 - Pe_{sf}) \, , \, R_{s0} = k_s^{te} * (1 - Pe_{s0})   \label{eq:Packet1}
\end{equation}
Assume $Pe_{sf} \leq Pe_{s0}$. Hence,  $R_{sf} \geq R_{f0}$.
Node $f$ acts as a forwarder of node $s$ and transmits for $k_f^{te}$ time slots.
During these time slots, $k_f^{te} * (1 - Pe_{f0})$ packets 
`successfully' reach the BS. Assume, 
\begin{equation}
R_f^{te} + R_c = k_f^{te} * (1 - Pe_{f0})   \label{eq:Packet3}
\end{equation}
Here, $R_f^{te}$ denotes the number of 'error-free' packets that contain
node $f$'s own data. $R_c$ represents the
number of `error-free' packets that contain node $s$'s data
and are forwarded by node $f$.

Let $R_s^{te}$ denote the goodput of node $s$ in TE.
Also assume that when node $s$ transmits,
the packets that get `lost' at node $f$ (the closer node), also get `lost' 
at node $0$ (the far node).
Based on this assumption and using the max-flow-min-cut theorem~\cite{Cover2},
\begin{equation}
R_s^{te} \leq \min \bigl(R_{sf}, R_{s0} + R_c \bigr) \label{eq:Packet_flow} 
\end{equation}

Based on this goodput analysis, we focus on the distributed joint
optimal time slot exchange and relay selection
in the sum goodput maximization of a TE network.

\section{Design Objective}  \label{sec:Theory}

\emph{Problem I}

\begin{subequations}
\begin{equation}
\mathbf{\max.}  \, \sum_{d \in \mathcal{D}} R_d^{te} +  \sum_{(s, \, f) \in \mathcal{SF}} 
\bigl(R_f^{te}  + R_s^{te})
 \label{eq:MainObjective}
\end{equation}
\begin{equation}
\mathbf{s.t.} \, \, (R_f^{te},R_s^{te}) \, \in \, conv(k_f^{te},k_s^{te}) 
\, \, \forall \, (s,f) \, \in \, \mathcal{SF}   \label{eq:RateConstraint1}  
\end{equation}
\begin{equation}
R_f^{te} \geq R_f^{in} \, , \, R_s^{te} \geq R_s^{in} 
\, \, \forall \, (s,f) \, \in \, \mathcal{SF}   \label{eq:RateConstraint2} 
\end{equation}
\begin{equation}
k_f^{te} + k_s^{te} \leq k_f^{in} + k_s^{in}, \, (k_f^{te}, k_s^{te}) \in \mathcal{Z}^+,  
\, \forall \, (s,f) \, \in \, \mathcal{SF}  \label{eq:BandwidthConstraint1}
\end{equation}
\begin{equation}
R_d^{te} = k_d^{te} * (1 - Pe_{d0}), \, k_d^{te} = k_d^{in}, \, \forall d \, \in \, \mathcal{D} 
 \label{eq:BandwidthConstraint2}
\end{equation}
\begin{equation}
\mathcal{D} \subseteq \mathcal{V} \, , \, \mathcal{SF} \in \mathcal{V} \times \mathcal{V} \, \, , \, \,
\mathcal{SF}_i \cap \mathcal{SF}_j = \emptyset  \, \, \forall i \neq j  \label{eq:RelaySelection1}
\end{equation}
\begin{equation}
\mathcal{SF}_i \cap \mathcal{D} = \emptyset \, \forall \, i \, \in [1,K]  
\label{eq:RelaySelection2}
\end{equation}
\begin{equation}
\mathcal{SF}_1 \cup \mathcal{SF}_2 \cdots \cup \mathcal{SF}_K \cup \mathcal{D} = \mathcal{V} 
\label{eq:RelaySelection3}
\end{equation}
\begin{equation}
Variables \, \, \mathcal{D}, \mathcal{SF}, R_s^{te}, R_f^{te}, k_s^{te}, k_f^{te}
\nonumber
\end{equation}
\end{subequations}

Equation~\eqref{eq:BandwidthConstraint2} and~\eqref{eq:Packet_Direct} suggest
that $R_d^{te} = R_d^{in}$. Therefore, the goodputs of the nodes in the direct set
$\mathcal{D}$ are not optimization variables. Equation~\eqref{eq:RateConstraint1} 
denotes that the goodputs of the forwarder and sender remain in the convex
hull of the allotted time slots, $k_s^{te}$ and $k_f^{te}$. This convex hull is
governed by~\eqref{eq:Packet1} and \eqref{eq:Packet3}. Equation~\eqref{eq:RateConstraint2} 
ensures that the goodputs of the sender and the forwarder through TE don't
drop below their initial goodputs. Equation~\eqref{eq:BandwidthConstraint1}
shows that the total time slots used by the sender and forwarder are 
constrained by the summation of the initial time slots allotted
to those nodes. Equation~\eqref{eq:RelaySelection1}-\eqref{eq:RelaySelection3}
denote that the direct node set and the sender-forwarder pairs cannot have
any common node and together, they form the overall set $\mathcal{V}$.

Problem I is similar to the design objective of our
earlier work on bandwidth exchange~\cite{Nazmul2}.
However, we focused on information theoretic capacity~\cite{Cover}
based resource allocation and relay selection in ~\cite{Nazmul2}. 
In this work, we focus on packet error probability based goodput
maximization, which is a tangible objective in an indoor wireless testbed.

The convex constraint of~\eqref{eq:Packet3},
the discrete time slot allocation and the sender-forwarder pair selection objectives
make problem I a mixed integer nonlinear programming (MINLP) problem. 
The solution of this MINLP involves an exponential number of 
variables and constraints. In the next section, 
we focus on designing a polynomial time distributed solution of problem I.

\section{Optimization Problem Solution}  \label{sec:Solution}

Let $R_{tot} = \sum_{i \in \mathcal{V}} R_i^{in}$ denote the 
summation of the initial goodputs of the nodes. 
For a fixed $\mathcal{SF}$, $R_{tot}$ can be
expressed in the following form:
\begin{eqnarray}
R_{tot} & = & \sum_{i \in \mathcal{V}} R_i^{in}   \nonumber \\
& = & \sum_{d \in \mathcal{D}} R_d^{in} +  \sum_{(s, \, f) \in \mathcal{SF}} 
\bigl(R_f^{in}  + R_s^{in} \bigr)  \label{eq:Modified1}   \\
& = & \sum_{d \in \mathcal{D}} R_d^{te} +  \sum_{(s, \, f) \in \mathcal{SF}} 
\bigl(R_f^{in}  + R_s^{in} \bigr)  \label{eq:Modified2} 
\end{eqnarray}
%
Equation~\eqref{eq:Modified2} uses the fact that $R_d^{te} = R_d^{in} \, \forall \, d \in D$.
Subtracting $R_{tot}$ from the objective function of I, we find
the following optimization problem:

\emph{Problem II}

\begin{subequations}
\begin{equation}
\mathbf{\max.}  \,  \sum_{(s, \, f) \in \mathcal{SF}} 
\bigl(R_f^{te}  + R_s^{te} - R_f^{in}  - R_s^{in})
 \label{eq:ModifiedObjective}
\end{equation}
\begin{equation}
\mathbf{s.t.} \, \, (R_f^{te},R_s^{te}) \, \in \, conv(k_f^{te},k_s^{te}) 
\, \, \forall \, (s,f) \, \in \, \mathcal{SF}
\label{eq:constraint1}  
\end{equation}
\begin{equation}
R_f^{te} \geq R_f^{in} \, , \, R_s^{te} \geq R_s^{in} 
\, \, \forall \, (s,f) \, \in \, \mathcal{SF}
\label{eq:constraint2}   
\end{equation}
\begin{equation}
k_f^{te} + k_s^{te} \leq k_f^{in} + k_s^{in}, \, (k_f^{te}, k_s^{te}) \in \mathcal{Z}^+,  
\, \forall \, (s,f) \, \in \, \mathcal{SF} 
\label{eq:constraint3}
\end{equation}
\begin{equation}
\mathcal{D} \subseteq \mathcal{V} \, , \, \mathcal{SF} \in \mathcal{V} \times \mathcal{V} \, \, , \, \,
\mathcal{SF}_i \cap \mathcal{SF}_j = \emptyset  \, \, \forall i \neq j 
\end{equation}
\begin{equation}
\mathcal{SF}_i \cap \mathcal{D} = \emptyset \, \forall \, i \, \in [1,K]  
\end{equation}
\begin{equation}
\mathcal{SF}_1 \cup \mathcal{SF}_2 \cdots \cup \mathcal{SF}_K \cup \mathcal{D} = \mathcal{V} 
\end{equation}
\begin{equation}
Variables \, \, \mathcal{D}, \mathcal{SF}, R_s^{te}, R_f^{te}, k_s^{te}, k_f^{te}
\nonumber
\end{equation}
\end{subequations}

The inclusion of constant terms does not change the optimal
variables of an optimization problem~\cite{Boyd}. Therefore, 
the optimal variables of both problem I and II are same. We focus
on solving problem II in the subsequent analysis and use the optimal
variables to find the solution of problem I.

Problem II involves both sender-forwarder pair selection and discrete time slot
allocation features. For a fixed set of sender-forwarder pairs,
the constraints in~\eqref{eq:constraint1}-~\eqref{eq:constraint3}
ensure that the discrete time slot exchange in one pair does not affect 
the other pairs. Hence, we now focus on an arbitrary sender-forwarder
pair $(s,f)$ and find the discrete time slot exchange in this pair. 

\subsection{Time Slot Allocation for a Fixed Sender
Forwarder Pair}

\emph{Problem III}

\begin{subequations}
\begin{equation}
\max \, \, \bigl(R_s^{te} - R_s^{in} \bigr) + \bigl(R_f^{te} - R_f^{in} \bigr)
 \label{eq:PacketSubObjective}
\end{equation}
\begin{equation}
R_{sf} = k_s^{te} * (1 - Pe_{sf}) \, , \, R_{s0} = k_s^{te} * (1 - Pe_{s0})
\label{eq:PacketSubConstraint1}
\end{equation}
\begin{equation}
R_f^{te} + R_c = k_f^{te} * (1 - Pe_{f0}) \, , \, 
R_s^{te} \leq \min \bigl(R_{sf}, R_{s0} + R_c \bigr)
\label{eq:PacketSubConstraint2}
\end{equation}
\begin{equation}
 R_f^{te} \geq R_f^{in} \, \, , \, \, R_s^{te} \geq R_s^{in}   
 \label{eq:PacketSubConstraint4}  
\end{equation}
\begin{equation}
k_f^{te} + k_s^{te} \, \leq \, k_f^{in} + k_s^{in} \, 
, \, (k_f^{te}, \, k_s^{te}) \, \in \mathcal{Z}^+ \, , 
\label{eq:PacketSubTimeConstraint}
\end{equation}
\begin{equation}
variables \, \,  
R_c, \, R_{sf}, \, R_{s0}, \, R_s^{te}, \, R_f^{te}, \, k_s^{te},
\, k_f^{te}  
\end{equation}
\end{subequations}

\underline{Lemma 2: }
Problem III is concave if $(k_f^{te},k_s^{te}) \in \mathcal{R}^+$.

\emph{Proof: } 
The objective function in~\eqref{eq:PacketSubObjective}
and the constraints in~\eqref{eq:PacketSubConstraint1}, \eqref{eq:PacketSubConstraint2}
and~\eqref{eq:PacketSubTimeConstraint} are linear.
Minimum of linear (concave) functions is concave~\cite{Boyd}. Hence,
the constraint in~\eqref{eq:PacketSubConstraint2} is convex. 
Thus, problem III is a concave maximization problem 
if $(k_f^{te},k_s^{te}) \in \mathcal{R}^+$. $\blacksquare$

The internal concave structure of problem III allows us to generate 
an upper and lower bound of problem III.

\subsubsection{Upper Bound} 

Let us modify problem III by relaxing the integer time slot constraint, i.e., let us assume
$(k_f^{te},k_s^{te}) \in \mathcal{R}^+$. Let's call it problem IV.
Now, the feasible region
of the modified problem is a superset of that of problem III.
Hence, the optimal solution of the modified problem is an upper 
bound of problem III. 
Denote this upper bound by $u^0$.

\subsubsection{Lower Bound} 
Let $k_s^{te,*}$ and $k_f^{te,*}$ denote the optimal time slot solutions of problem IV. 
 Now, if $k_s^{te,*}$ and $k_f^{te,*}$ are integers, they are also the
 optimal solutions of problem III. 
Otherwise, convert $k_s^{te,*}$ and $k_f^{te,*}$ to the nearest integers and find the 
corresponding goodputs. If the corresponding solution 
is feasible for problem III, it can serve as a lower bound to 
the optimal solution of problem III.
Denote this lower bound by $l^0$.

\subsubsection{Computational Complexity of Problem III}

Probelm III can be optimally solved by searching over $k_s^{in}$ time
slot transfers. However, one can further reduce the complexity
by solving the convex programming based upper and lower
bounds of problem III. Let $\epsilon$ denote the tolerance of 
the optimal solution. If $(u_0 - l_0) \leq \epsilon$, we can use the lower bound as the solution
of problem III.

\subsection{Optimal Sender Forwarder Pair Selection}

The optimal solution of problem III denotes the 
sender-forwarder pair's goodput gain through cooperation, 
over non-cooperation. The optimal relay selection part of problem II is to find the 
set of sender-forwarder pairs that maximizes the summation of the goodput gain. 
Now, consider a graph $ \mathcal{G} = (\mathcal{V},\mathcal{E})$ 
where the vertices $\mathcal{V}$ represent the set of $N$ nodes 
under consideration and $\mathcal{E}$ denotes the edges between these nodes.
Define the edge weight of any $(i,j)$ pair by 
$R_i^{te} + R_j^{te} - R_i^{in} - R_j^{in}$, i.e., the difference, in terms of goodput, 
between the cooperation and non-cooperation scenario.
Using the interference-free scheduling algorithm
of~\cite{Ephremides}, the optimal sender-forwarder
set selection problem can be shown to be equivalent to solving
the maximum weighted matching (MWM) algorithm~\cite{Edmonds}
in the above graph. 
A detailed proof of the equivalence between the relay selection
problem and the MWM algorithm can be found in~\cite{Nazmul2}.
The next section will illustrate the use of MWM in the 
optimal sender-forwarder pair selection among $3$ testbed nodes.

The MWM algorithm can be distributively solved
using the distributed local greedy MWM~\cite{Preis}.
Distributed MWM finds the pairs by selecting the
locally heaviest edges and guarantees at least $50\%$ performance
of the optimal solution of problem I~\cite{Preis}. Our distributed TE implementation
protocol is based on the distributed MWM and will be described in the next section.

\subsection{Distributed TE Protocol} 

\begin{figure}[t]
\centering
\includegraphics[scale=0.29]{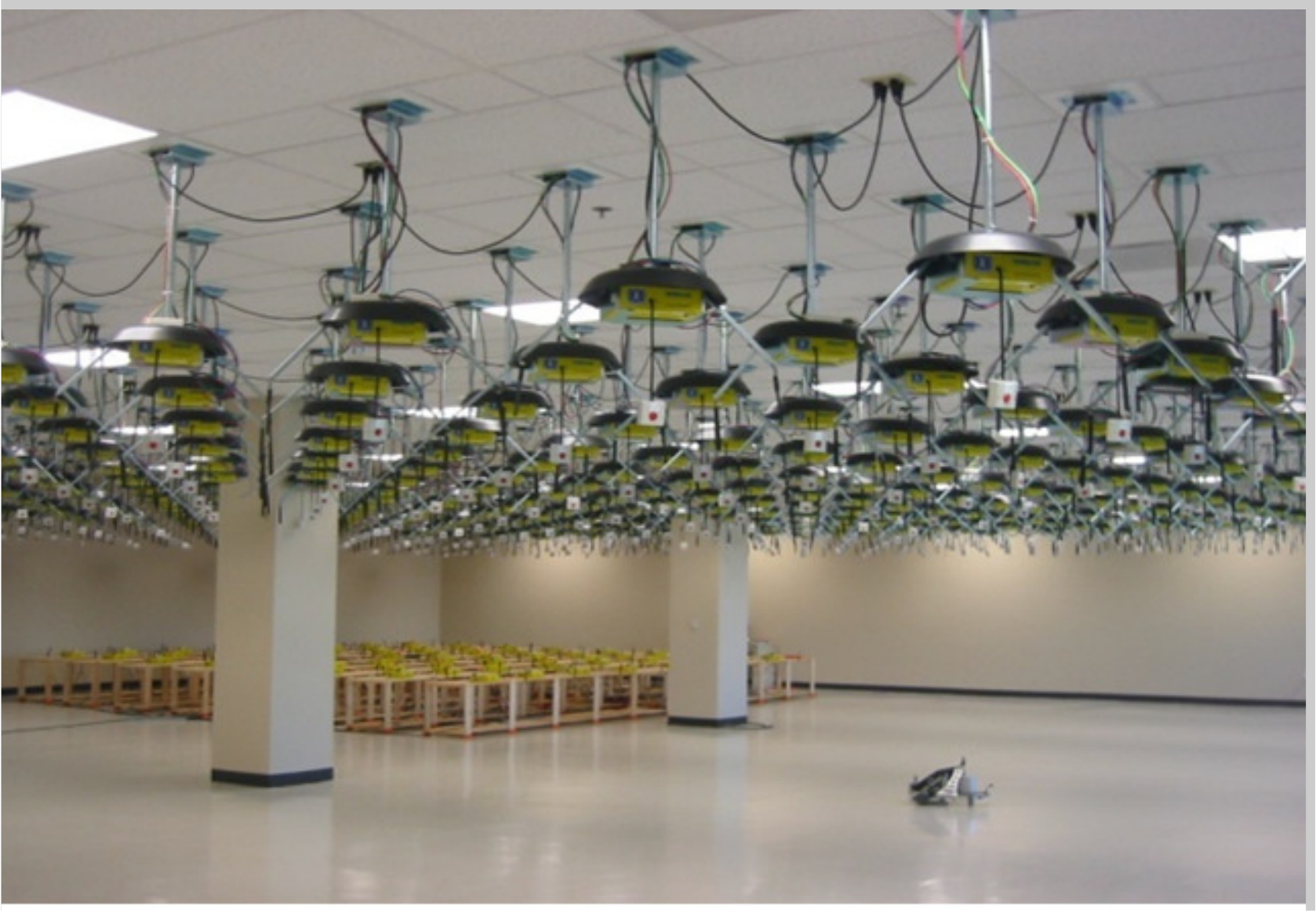}
\caption{Orbit Testbed}
\centering
\label{fig:Orbit}
\end{figure}

\begin{enumerate}

\item Node $i \, \in \, \mathcal{V}$ has
$k_i^{in}$ transmission time slots and knows the packet error probability of its direct path, $Pe_{i0}$,

\item Let $\mathcal{J}$ be the set of neighbours of $i$. 
In a wireless environment, neighbouring nodes can hear each other.
Node $j \, \in \, \mathcal{J}$
receives node $i$'s packets and can calculate the inter-node
packet error probability, $Pe_{ij}$.

\item Node $i$ sends an omnidirectional message containing 
$Pe_{i0}$ and $k_i^{in}$ to its neighbours.


\item Node $i$ solves problem III for all $j \, \in \, \mathcal{J}$
and finds the goodput gain for each of its neighbours.
Thus, node $i$ knows its adjacent link weights.

\item Node $i$ picks the ``candidate" node $j$, based on the heaviest
adjacent link weight and sends ``add" request.

\item If node $i$ receives an ``add" request from node $j$, $i$ and $j$
form a cooperative pair. Node $i$ sends ``drops" request to its other neighbours.

\item If node $i$ receives a ``drop" request from $j$, node $i$ removes
the $(i,j)$ link from its adjacent edge set. Node $i$ returns to step $5$.

\item The pair selection process converges after at most $N^2$ message passings.
The `matched' nodes form the set of cooperative pairs $\mathcal{SF}$.
The `unmatched' nodes form $\mathcal{D}$ and transmit data without
cooperation.

\item The sender node $s$ of a cooperative pair transmits its 
own data during $k_s^{te}$ time slots. 

\item The BS sends ACK of the `correctly' received packets to the sender node.
The forwarder node also hears
these ACK messages. Based on this information, node $f$ 
finds the packets of $s$ that got `lost' at the BS.

\item Node $f$ transmits for $k_f^{te}$ time slots. During these
slots, node $f$ forwards the `lost' packets of $s$ and then 
transmits its own data packets to the BS.

\end{enumerate}

\subsection{Computational Complexity of the Distributed Algorithm}

Each node $i \, \in \, \mathcal{V}$ solves problem III
for each of its neighbours. Let $n_i$ be the number of neighbours
of node $i$. Since problem III can be optimally solved using $O(k_i^{in})$
searches, each node performs $O(n_i k_i^{in})$ computations.
The total number of computations in the $N$ node network
is $O(n_i k_i^{in} N)$. 
 
The number of time slots allotted to node $i$, $k_i^{in}$, can be
approximated as, $k_i^{in} \approx \frac{k^{tot}}{N}$.
Since $n_i << k^{tot}$, the overall complexity is,
$O(k^{tot})$.

Thus, the proposed TE protocol solves an approximated
distributed version of problem I with $O(k^{tot})$
complexity and at most $N^2$ message passings.

\section{Experimental Setup}  \label{sec:Setup}

\subsection{ORBIT Testbed \& USRP Nodes}

\begin{figure}[t]
\centering
\includegraphics[scale=0.31]{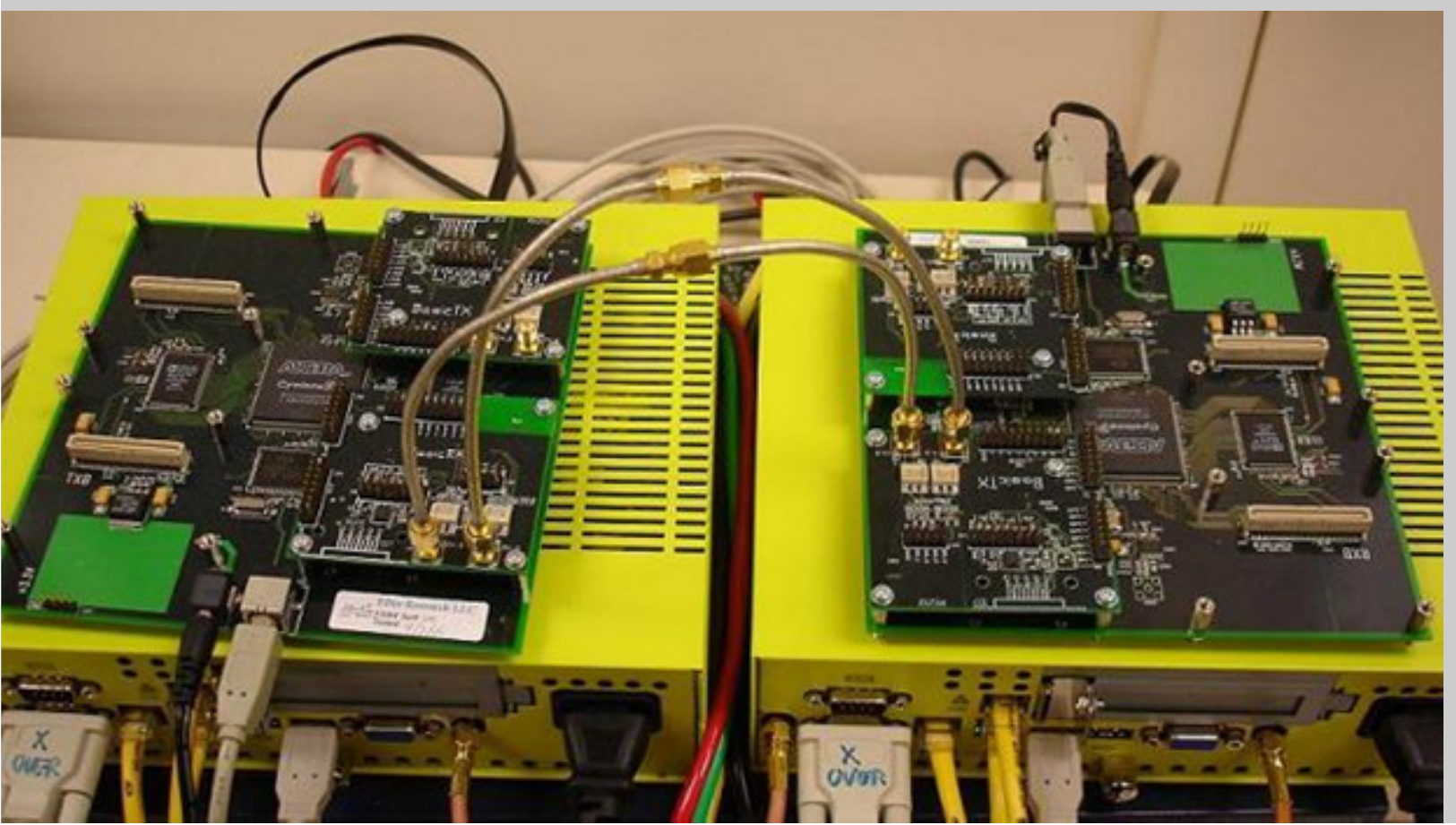}
\caption{USRP Daughterboards}
\centering
\label{fig:USRP}
\end{figure}

We implement the proposed TE based incentivized algorithm
among the USRP nodes of ORBIT, an indoor wireless testbed of 
Wireless Information Network Laboratory (WINLAB), Rutgers University. 
ORBIT has $400$ nodes, overall, in a $20m \times 20m$ square grid. 
Fig.~\ref{fig:Orbit} shows a snapshot of the ORBIT testbed. 

ORBIT has $15$ USRP nodes that can be used in software defined
radio based experiments. Fig.~\ref{fig:USRP} shows the snapshots
of two USRP daughter boards. We use the GNU radio software toolkil~\cite{GNUradio}
to run experiments in these USRP nodes. Specifically,
we use the benchmark-tx.py and benchmark-rx.py codes to transmit
and receive packets between two USRP nodes~\cite{GNUradio}. 
The flexibility of GNUradio allows us to change the transmission 
power level and packet sizes through software. 
This variable power capability of GNUradio,
along with the spatial separation among the nodes, allow
us to create links with different strengths between
different node pairs. 

\begin{figure}[t]
\centering
\includegraphics[scale=0.35]{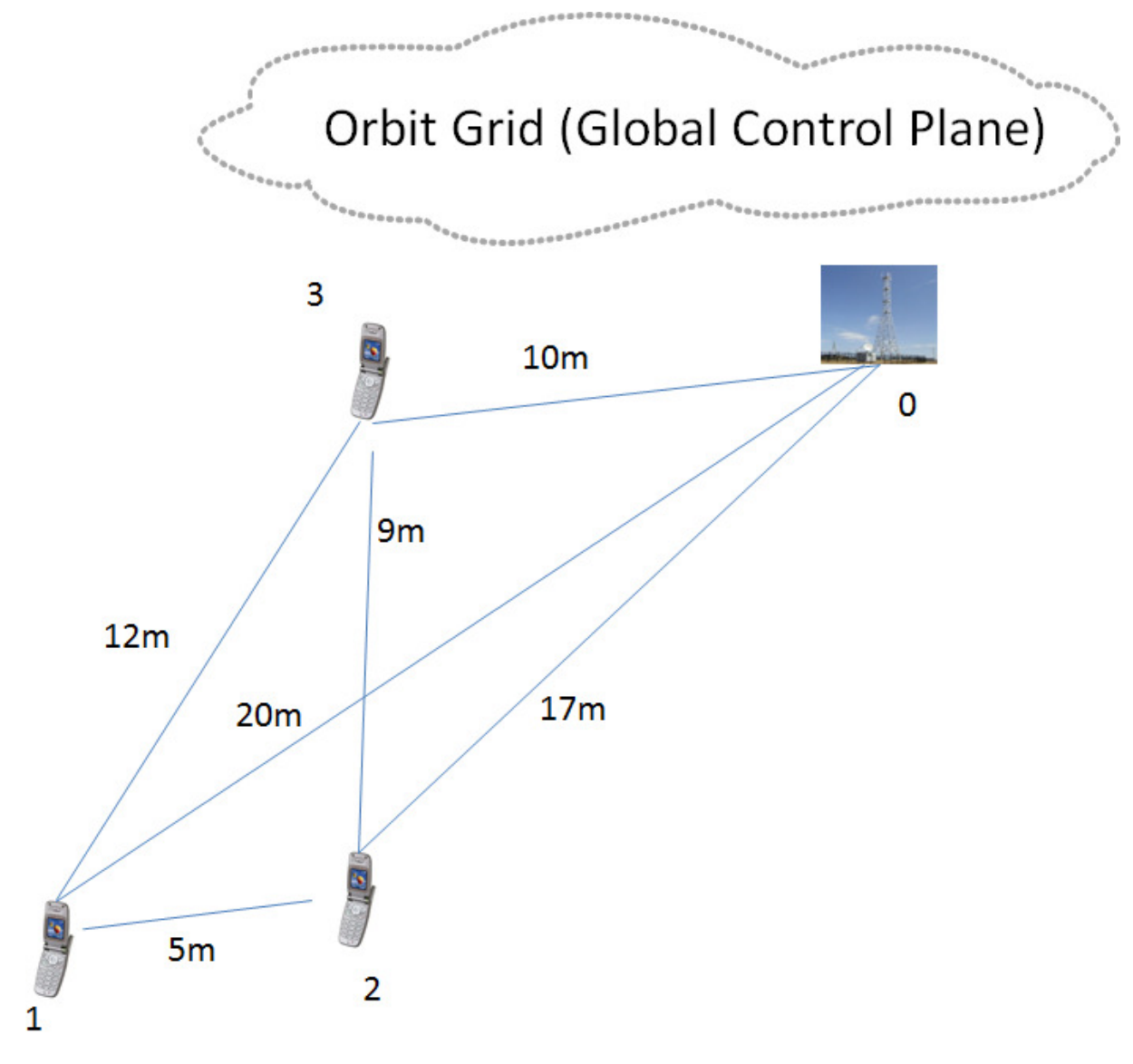}
\caption{Spatial Separation of Selected Nodes}
\centering
\label{fig:node_distance}
\end{figure}

As shown in Fig.~\ref{fig:node_distance}, we use four USRP 
nodes of the ORBIT testbed to conduct the TE based cooperative forwarding experiments.
Fig.~\ref{fig:node_distance} also shows the
spatial separation of the selected nodes. 
Here, node $1$, $2$ and $3$
constitute the user set $\mathcal{V}$ and node $0$ serves as the BS.
The ORBIT grid is used as a global control plane
to exchange the control information between the nodes.

\subsection{Selection of Parameters}   

The benchmark-tx.py and benchmark-rx.py codes of GNUradio
allow the following four modulation schemes:
a) GMSK, b) differential binary phase shift keying (DBPSK),
c) differential quadrature phase shift keying (DQPSK) and
d) differential 8 phase shift keying (D8PSK). 
DBPSK, DQPSK and D8PSK are found to be very sensitive to peak power clippings
due to their variable envelope waveform.
Therefore, we use a fixed modulation scheme, GMSK, in our experiments. 

Each node transmits at $1$ Mbps and each packet contains $1500$ bytes. As a result, it
takes $(1500 * 8)/(1 * 10^6)$, i.e., 0.012 second
to transmit one packet. We assume each time slot to
be $0.012$ second long, i.e., one packet is transmitted in each slot.
 
The total transmission time is assumed to be $3$ second. Each node 
is initially allotted $1$ second transmission time, i.e., 
$1/0.012$ or $83$ time slots. We approximate the number of time 
slots since fractional packet transmission is not considered.

We also add $32$ bit CRC sequence in each packet
and make it similar to the Ethernet packet structure~\cite{Ethernet}.
Note that, we do not use error control coding in these experiments.
Therefore, the presence of a single bit error leads to the `loss'
of the whole packet due to CRC.

\section{Experimental Evaluation}  \label{sec:Results}

\subsection{Illustration of MWM in Relay Selection}

\begin{figure}[t]
\centering
\includegraphics[scale=0.34]{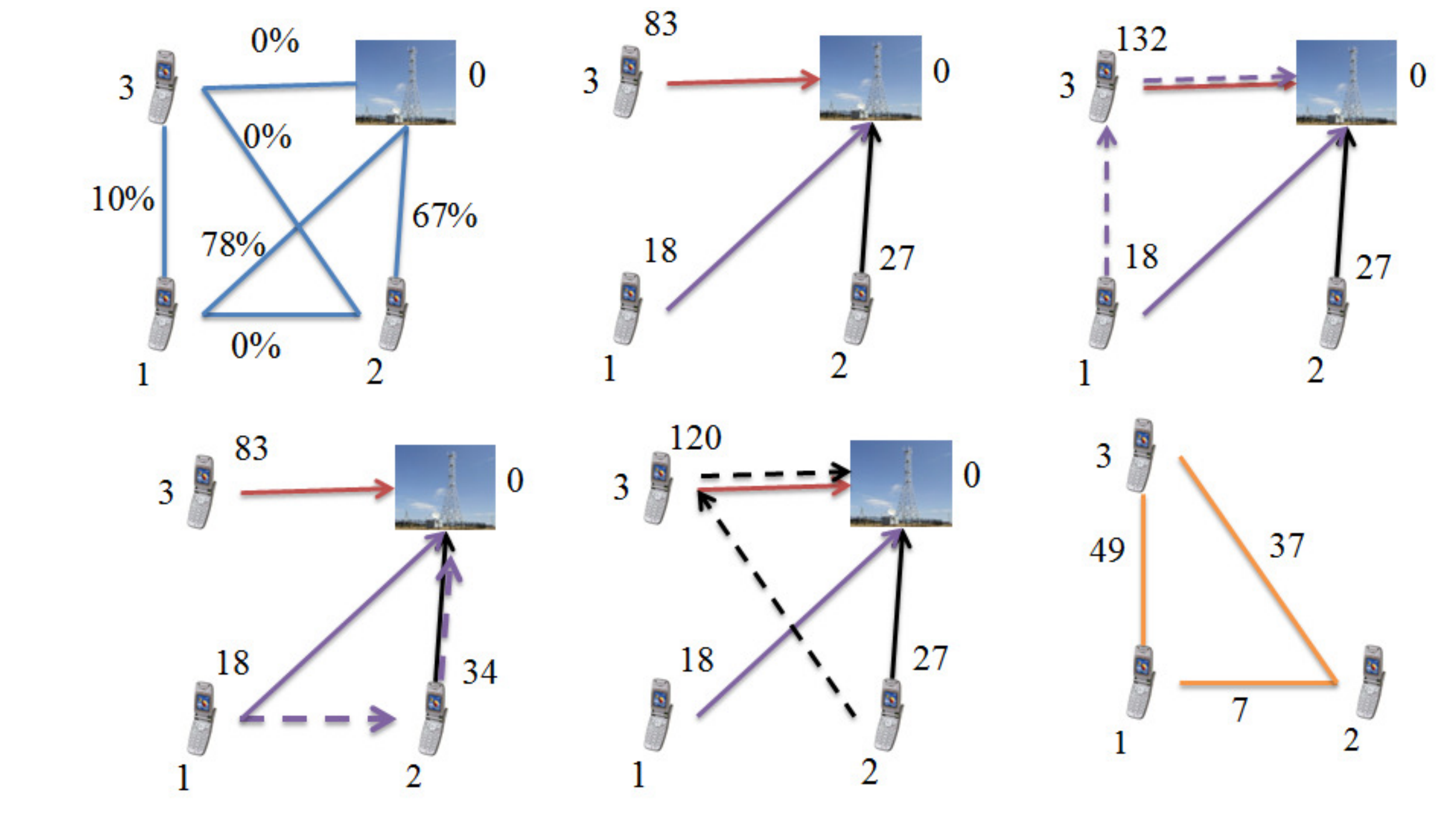}
\caption{Illustration of MWM in sender-forwarder pair selection}
\centering
\label{fig:MWM}
\end{figure}

Fig.~\ref{fig:MWM} shows the use of MWM in the optimal
sender-forwarder pair selection among $3$ testbed nodes.
The left figure of the top row shows the packet loss probability between the
inter-node pairs. These packet error probabilities were based on $1500$ byte packet 
length, GMSK modulation, some fixed power level and CRC checking. 
The middle figure of the top row focuses on the direct transmission 
scenario and shows the goodput (in packet/3 second) of each node. 
Each node initially receives
$83$ time slots and transmits one packet at each slot through the direct path.
The packet error probability in link $30$ is $0\%$. Therefore, all transmitted
packets of node $3$ reach the BS. Node $1$ and $2$'s goodputs are considerably lower
due to the high packet error probability in link $10$ and $20$ respectively.

The top right, the bottom left and the bottom middle figure
show the goodput (in packet/3 second) of
different sender-forwarder cooperation scenarios. The top right figure 
focuses on the TE based cooperation between node $1$ and $3$.
Here, node $1$ and $3$ solve the two node time slot allocation optimization
of problem III. The cooperation allows node $3$ to achieve a goodput
of $132$ packets and ensures that node $1$'s
goodput does not drop below $18$ packets, its initial value.
Therefore, the overall goodput gain obtained through the cooperation of node $1$ and $3$ 
is $49$ packets. As a result, the $13$ link of the MWM graph,
shown in the bottom right figure, is assigned a weight of $49$.

The bottom left and bottom middle figures demonstrate the cooperation
scenario in node 1--2 and 2--3 respectively. The bottom right figure shows
the link weight of the corresponding cooperation pairs. 
The distributed local greedy MWM selects link $13$. 
Therefore, node $1$ and $3$ cooperate using TE, whereas, 
node $2$ transmits without cooperation.

\begin{figure}[t]
\centering
\includegraphics[scale=0.40]{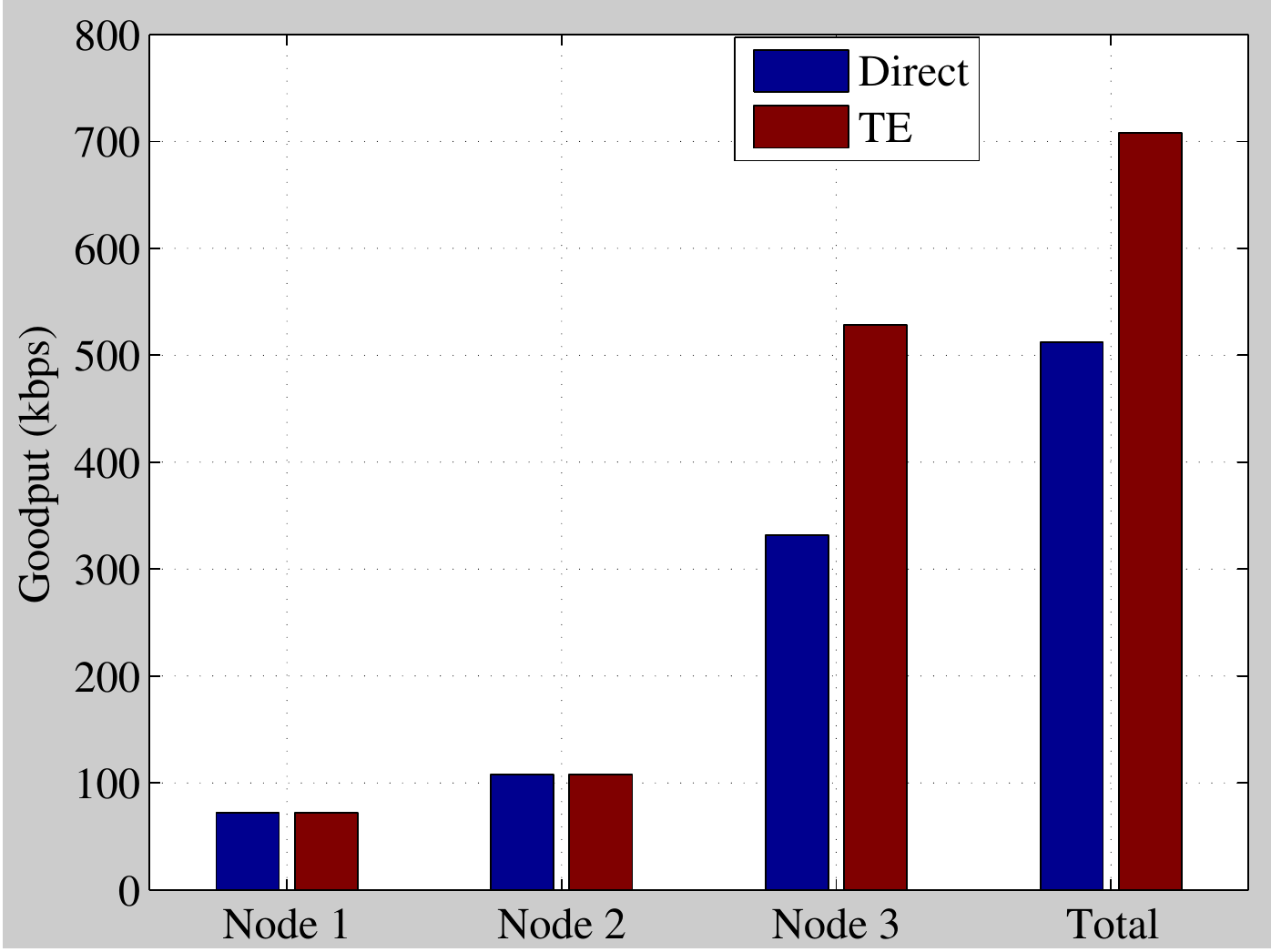}
\caption{Sum Goodput Maximization in 3 node 
(Packet length = 1500 bytes, CRC checking, GMSK modulation)}
\centering
\label{fig:Sumrate}
\end{figure}

\subsection{Sum Goodput Maximization} 

Fig.~\ref{fig:Sumrate} compares the sum goodput (in kilo bit per second (kbps))
of TE and direct path transmission.
Fig.~\ref{fig:MWM} shows that node $1$ and $3$ get 
selected as the cooperative pair due to the MWM algorithm.
Therefore, node $1$ and $3$ solve problem III to find the optimal
time slot transfers. Due to the sum goodput maximization objective,
the benefits of cooperation go to node 3, i.e., the node with the better channel. 
Node $3$'s goodput increases by $70\%$. 
The constraint of~\eqref{eq:PacketSubConstraint4} ensures that
node $1$ gets  its initial goodput, at least. On the other hand, 
node $2$ transmits without cooperation 
and its goodput does not change from the initial value.

\subsection{Proportional Fair Maximization of Goodput}

\begin{figure}[t]
\centering
\includegraphics[scale=0.40]{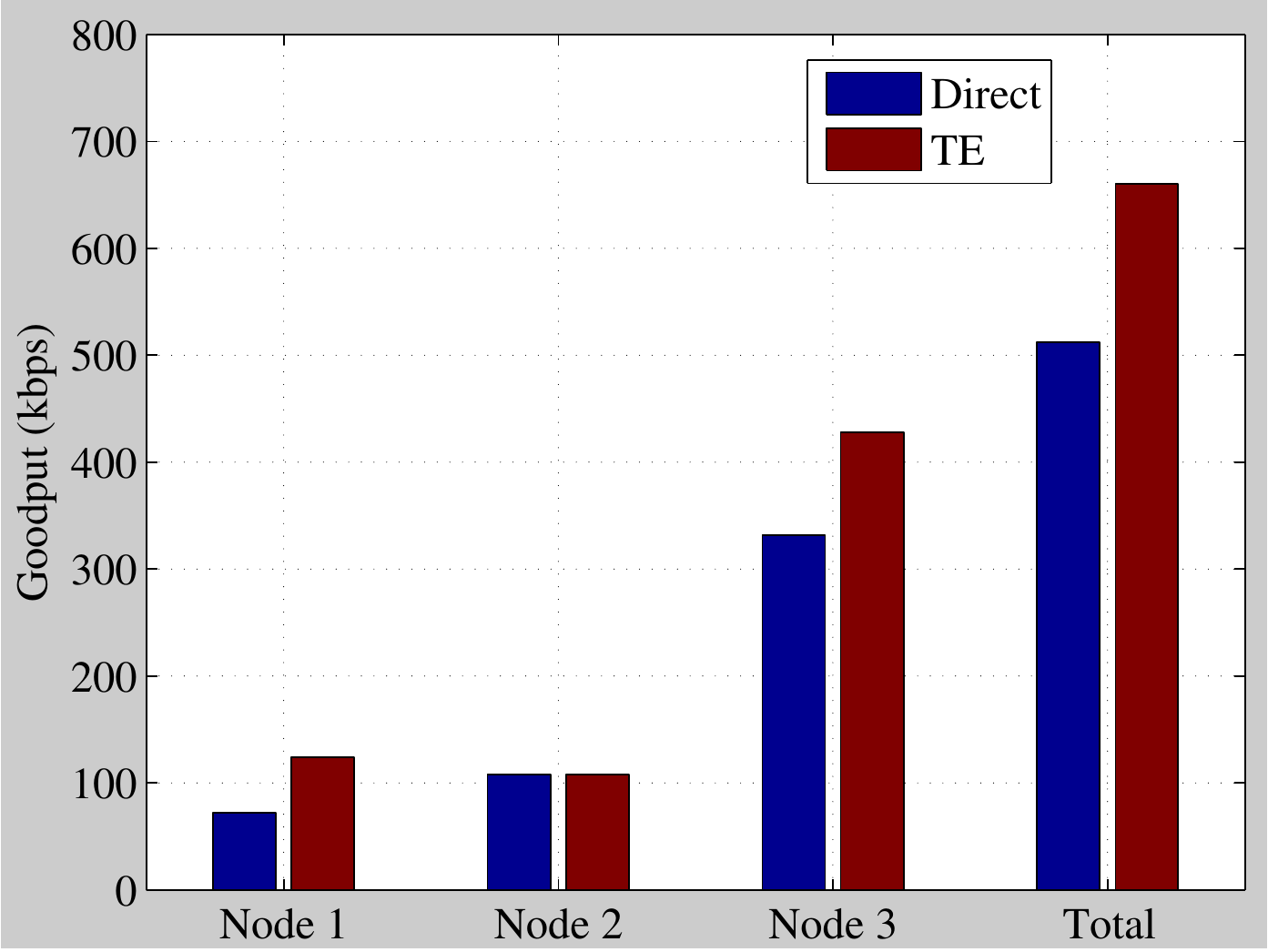}
\caption{Proportional Fair Maximization of Goodput
(Packet length = 1500 bytes, CRC checking, GMSK modulation)}
\centering
\label{fig:Fair}
\end{figure}

Fig.~\ref{fig:Fair} compares the proportional fair maximization
performance of direct transmission and TE. Here, the 
selected cooperative nodes, $s$ and
$f$, solve a modified version of problem III. 
In this modified problem, $s$ and $f$ maximize $(R_s^{te} - R_s^{in}) * (R_f^{te} - R_f^{in})$
instead of maximizing $(R_s^{te} + R_f^{te})$. Hence,
the goodput of both nodes increase due to cooperation.
Fig.~\ref{fig:Fair} shows that the goodputs of node $1$ and $3$
increase by $70\%$ and $30\%$ respectively.

\section{Conclusion}  \label{sec:Conclusion}

We designed and implemented TE based cooperative forwarding
among the USRP nodes in the ORBIT indoor wireless testbed. 
We solved the joint time slot allocation and
sender-forwarder pair selection problem in this setup. 
Our proposed algorithm maximizes the global goodput of the
network while ensuring that no node's goodput drops below its initial value. 
The ORBIT grid is used as a global control plane to exchange the control information
between the USRP nodes.
Experimental results suggest that resource delegation based 
cooperative forwarding can significantly improve the sum goodput
and proportional fair goodput performance of the network.

The use of adaptive modulation
and signal to noise ratio based resource allocation in testbed implementation
remains an area of future research.

\section{Acknowledgements}

This work is supported by the Office of Naval Research under grant N00014-11-1-0132.
We thank Kush Patel, Sid Paradkar and Hakim Ergaibi for their assistance in GNUradio coding
and testbed implementation. 

\bibliographystyle{IEEEbib}
\bibliography{Globecomm_bibtex}

\end{document}